\documentclass[lettersize,journal]{IEEEtran}
\usepackage[frozencache,cachedir=.]{minted} 
\usepackage{amsmath,amsfonts}
\usepackage{algorithmic}
\usepackage{algorithm}
\usepackage{array}
\usepackage[caption=false,font=normalsize,labelfont=sf,textfont=sf]{subfig}
\usepackage{textcomp}
\usepackage{stfloats}
\usepackage{url}
\usepackage{verbatim}
\usepackage{graphicx}
\usepackage{cite}
\usepackage{caption}
\usepackage{enumitem,kantlipsum}
\usepackage[usenames,dvipsnames]{xcolor}
\usepackage{colortbl}
\usepackage[all]{tcolorbox}
\usepackage{tabularx}
\usepackage{stfloats}
\usepackage{pifont}
\usepackage{url}
\usepackage{verbatim}
\usepackage{graphicx}
\usepackage{caption}
\usepackage{pifont}
\newcommand{\xmark}{\ding{55}}

\usepackage{color}
\usepackage{listings}
\definecolor{codegreen}{rgb}{0,0.6,0}
\definecolor{codegray}{rgb}{0.5,0.5,0.5}
\definecolor{codepurple}{rgb}{0.58,0,0.82}
\definecolor{backcolour}{rgb}{0.95,0.95,0.92}

\lstdefinestyle{mystyle}{
    backgroundcolor=\color{backcolour},   
    commentstyle=\color{codegreen},
    keywordstyle=\color{magenta},
    numberstyle=\tiny\color{codegray},
    stringstyle=\color{codepurple},
    basicstyle=\ttfamily\footnotesize,
    breakatwhitespace=false,         
    breaklines=true,                 
    captionpos=b,                    
    keepspaces=true,                 
    numbers=left,                    
    numbersep=5pt,                  
    showspaces=false,                
    showstringspaces=false,
    showtabs=false,                  
    tabsize=2
}

\tcbset{tab2/.style={enhanced,fonttitle=\bfseries, fontupper=\small\sffamily,
colback=blue!5!white,colframe=blue!50!black,colbacktitle=cyan!40!white,
coltitle=black,center title}}

\begin{document}

\title{Bluetooth and WiFi Dataset for Real World RF Fingerprinting of Commercial Devices}

\author{\IEEEauthorblockN{Anu Jagannath${^{\ddagger\dagger}}$, \textit{Senior Member}, IEEE, Zackary Kane$^\ddagger$, and Jithin Jagannath${^{\ddagger \aleph}}$, \textit{Senior Member}, IEEE, 
}


\IEEEauthorblockA{
$^{\ddagger}$Marconi-Rosenblatt AI/ML Innovation Lab, ANDRO Computational Solutions, LLC, Rome NY \\
$^{\dagger}$Institute of Wireless Internet of Things, Northeastern University, Boston, MA \\ 
$^{\aleph}$ University at Buffalo, Buffalo, NY,  Email: \{ajagannath, zkane, jjagannath\}@androcs.com
}}

\markboth{Journal of \LaTeX\ Class Files,~Vol.~14, No.~8, August~2021}%
{Shell \MakeLowercase{\textit{et al.}}: A Sample Article Using IEEEtran.cls for IEEE Journals}


\maketitle

\begin{abstract}
 RF fingerprinting is emerging as a physical layer security scheme to identify illegitimate and/or unauthorized emitters sharing the RF spectrum. However, due to the lack of publicly accessible real-world datasets, most research focuses on generating synthetic waveforms with software-defined radios (SDRs) which are not suited for practical deployment settings. On the other hand, the limited datasets that are available focus only on chipsets that generate only one kind of waveform. Commercial off-the-shelf (COTS) combo chipsets that support two wireless standards (for example WiFi and Bluetooth) over a shared dual-band antenna such as those found in laptops, adapters, wireless chargers, Raspberry Pis, among others, are becoming ubiquitous in the IoT realm. Hence, to keep up with the modern IoT environment, there is a pressing need for real-world open datasets capturing emissions from these combo chipsets transmitting heterogeneous communication protocols.  
To this end, we capture the first known emissions from the COTS IoT chipsets transmitting WiFi and Bluetooth under two different time frames. The different time frames are essential to rigorously evaluate the generalization capability of the models. To ensure widespread use, each capture within the comprehensive 72 GB dataset is long enough (40 MSamples) to support diverse input tensor lengths and formats. Finally, the dataset also comprises emissions at varying signal powers to account for the feeble to high signal strength emissions as encountered in a real-world setting. 
\end{abstract}


\section{Introduction}
\IEEEPARstart{T}{he} unprecedented adoption of Internet of Things (IoT) devices in everyday life especially WiFi and Bluetooth have exacerbated privacy and security concerns. Radio Frequency (RF) fingerprinting is envisioned as a physical layer security scheme that can be integrated with the existing upper layer security protocols to ensure robust wireless security \cite{device_fingerprint,AJagannathComnet22}. RF fingerprinting is a passive security scheme that can be implemented at a passive receiver radio whereby the wireless emitters can be distinguished from one another based on the minute imprint made in its emissions due to the imperfections in its RF circuitry. Such minute RF circuitry signatures can be identified even when they belong to the same manufacturer and emit the same waveform. 

In recent years, there has been immense interest among the research community to study RF fingerprinting of different types of emitters such as WiFi, Bluetooth, LoRa, and ADS-B, among others \cite{lora_rff_shen,wifi_indoor_survey}. There has been a lot of research regarding the application of deep neural networks (DNN) in generalizing and effectively distinguishing the various RF emitters. A key factor in promoting active deep learning (DL) based research in RF fingerprinting is the availability of open RF datasets corresponding to the most common wireless standards such as WiFi, Bluetooth, LoRa, Zigbee, etc., transmitted from IoT devices. Most openly available datasets are either synthetically generated and emitted with software-defined radios (SDRs) or same time frame (i.e., over the course of under a month) \cite{alshawabka,trust5G} captures consequently rendering it hard to perform and develop any real-world RF fingerprinting approaches tailored towards practical deployment. In this article, we, therefore, share with the research community a large RF dataset (72 GB) comprising WiFi as well as Bluetooth emissions from 10 commercial off-the-shelf (COTS) IoT chipsets captured over multiple days\cite{AJ_BT_WIFI_IoT}. The captures over two different time frames will enable conducting a generalization test to evaluate the trained DNN model's ability in identifying emitters under vastly different propagation, interference, and multipath scenario than seen with the training dataset. Such generalization tests are critical in assessing the DNN's performance under a practical deployment setting. Therefore, this dataset is envisioned to immensely benefit the research community in actively fostering and developing RF fingerprinting approaches tailored toward real-world deployment.


\section{Related Datasets}

Despite the active research in RF fingerprinting, there is an existing gap in the availability of heterogeneous real-world RF datasets. For example, the extensive RF fingerprinting dataset developed as part of the Defense  Advanced  Research Projects Agency (DARPA) radio frequency machine learning systems (RFMLS) program \cite{rfmls} is not available to the public. The RFMLS dataset contains emissions from 5117 WiFi and 5000 ADS-B emitters captured over multiple days. The authors of \cite{alshawabka} released a WiFi dataset from 20 transmitters comprising captures spanning 10 days under different antenna and other capture scenarios. However, the WiFi waveforms are synthetically generated and transmitted with 20 universal software  radio peripheral (USRP) SDRs. RF fingerprinting of 4 base station emitters transmitting 5G New Radio (NR), WiFi, and LTE were studied in \cite{trust5G}. Here again, the waveforms are synthetically generated using MATLAB WLAN, LTE, and 5G toolboxes and transmitted from USRP radios. However, such synthetically generated waveforms emitted using SDRs are not representative of the hardware signatures from COTS radios. 

A narrowband (2 MHz) Bluetooth dataset from 10 COTS emitters spanning captures from multiple days 
was adopted for an embedding-based attentional RF fingerprinting solution in \cite{AJagannathBTEmbedATN}. However, in this article we release a wideband (66.67 MHz) capture of both WiFi as well as Bluetooth waveforms from the same 10 COTS emitters in the IEEE dataport \cite{AJ_BT_WIFI_IoT}. A large-scale WiFi signal dataset from 174 COTS emitters captured over a span of 1 month is described in \cite{wisig}. 

\begin{table*}
\centering
\caption{Comparison with state-of-the-art.} \label{tab:compare}
\begin{tcolorbox}[width=17.5 cm,tab2,tabularx={p{3.5 cm}|p{2 cm}|p{2.5 cm}|p{1.5 cm}|p{2 cm}|p{1.5 cm}|p{1.5 cm}}]
\textbf{Dataset} & \textbf{Real-world or \newline synthetic} & \textbf{Supported Waveforms} & \textbf{Openly Available}& \textbf{Aging Considered} & \textbf{Varying Locations} & \textbf{Varying Power} \\
\hline
Proposed work &Real-world &Bluetooth, WiFi & \checkmark & \checkmark & \checkmark & \checkmark\\
DARPA RFMLS \cite{rfmls} &Real-world &ADS-B, WiFi & \xmark & \checkmark & \xmark & \checkmark\\
Al-Shawabka et al. \cite{alshawabka} &Synthetic &WiFi &\checkmark & \xmark & \checkmark & \xmark\\
Reus-Muns et al.\cite{trust5G} &Synthetic &LTE, 5G-NR, WiFi &\checkmark & \xmark & \xmark & \xmark\\
Jagannath et al. \cite{AJagannathBTEmbedATN} & Real-world &Bluetooth &\checkmark & \checkmark & \checkmark & \checkmark\\
Hanna et al. \cite{wisig} &Real-world &WiFi &\checkmark & \checkmark & \xmark & \xmark\\\hline
\end{tcolorbox}
\end{table*}
In contrast, the dataset introduced in this article possesses emissions from 10 COTS emitters over the span of several months and comprises WiFi as well as Bluetooth in an effort to study the impact of diverse wireless standards from the same emitter. Our main contributions can be summarized as follows:
 \begin{enumerate}
    \item We provide the research community access to emissions from commercial hardware collected under real-world settings to allow them to evaluate the real-world performance of their RF fingerprinting solutions. We argue that the artificially simulated and introduced hardware distortions are not representative of the manufacturing defects and commercial waveform standards emitted from such actual IoT emitters subsequently thwarting the ability to showcase the actual performance of AI/ML algorithms deployed on real-world commercial hardware. 
     \item This released dataset \cite{AJ_BT_WIFI_IoT} provides the raw version of the captured IQ samples instead of the preprocessed samples as in \cite{wisig} and of very large length (40 MSamples) allowing the flexibility to adopt any user-defined preprocessing methodologies and sample lengths to suit their AI/ML algorithm. For example, some AI/ML algorithms would require the use of hand-crafted features such as magnitude, phase, Power Spectrum Density (PSD), Wigner-Ville distributions, or other time-frequency transformations to be performed on the training samples while others would benefit from raw IQ samples. Such hand-crafted features can only be extracted from raw unprocessed IQ samples as opposed to samples of finite length pre-processed by some predefined functions. Hence, we ensure the release of raw IQ samples for the diverse and extended use of the dataset for the wider research community.
     \item This dataset will enable the research community to study the impact of diverse wireless standards emitted from the same emitter. In other words, this will ensure that the RF fingerprinting algorithm identifies the emitter accurately rather than misinterpreting it as a different emitter when transmitting a different wireless standard than the one trained on. Combo chipsets are commonly found in most IoT devices such as laptops, and mobile phones, among others emphasizing the practical relevance of training the RF fingerprinting algorithms with such diverse waveforms transmitted from the same emitter. However, there are no available datasets that mitigate this gap. All of the available datasets include only singular wireless standards (WiFi, LoRa, or Zigbee, among others). Further, the benefit of using multi-task learning has been established in \cite{Baxter1997} where it was shown that it reduces the risk of overfitting by the order of the number of tasks. This has been demonstrated in the realm of modulation classification in \cite{AJagannath22PHYCOM}. To aid multi-task learning, the annotations facilitating it such as wireless standard, emitter identifier, and modulation among others should be available in the dataset. Therefore, in this dataset, we have provided all the necessary annotations in the metadata to enable such multi-task applications as well. This becomes even more relevant as the adoption of such combo chipsets is becoming widespread. Hence, we aim to address this gap by providing the research community with actual emissions from combo chipsets.  
     \item We further partition the dataset based on their time frame of collection to enable the user to evaluate the generalization capability of their RF fingerprinting solutions. The device fingerprint is influenced by several confounding factors such as the circuitry aging, and propagation channel, among others. The propagation channel in an uncontrolled indoor laboratory will vary significantly on a day-to-day basis. Similarly, the circuitry will suffer aging due to continuous use and handling. Practical deployments will involve several such confounding factors and are intended to work over prolonged periods of time. Such nuances cannot be realistically simulated in synthetically generated data samples. To facilitate this, we provide data samples that were collected during different time frames. We segment these data samples that were separated by few months time period and notate it for ease as Day-1 and Day-2 within each WiFi and Bluetooth datasets.
 \end{enumerate}
 We have also denoted the advantages and features of the proposed dataset with other state-of-the-art RF fingerprinting datasets in Table \ref{tab:compare}. 
\section{Setup of IoT Devices: Dataset Construction}
\subsection{Testbed Layout}

\begin{figure}[h!]
         \centering
         \includegraphics[angle=-90, width=0.9 \columnwidth]{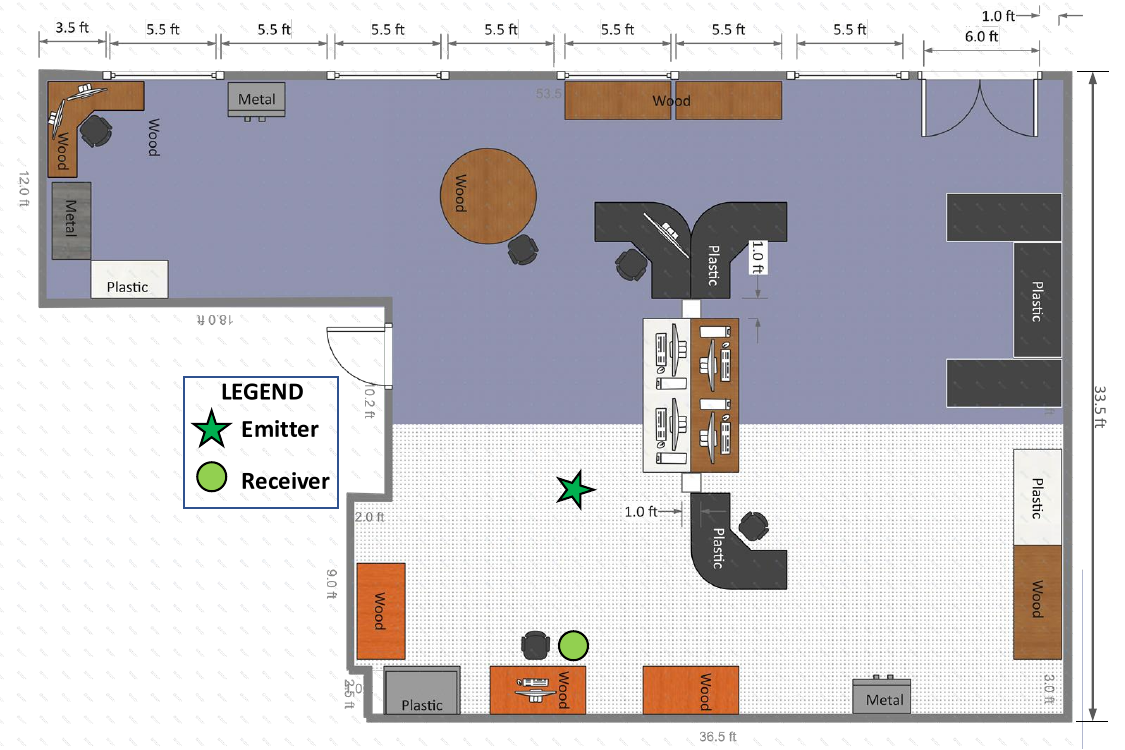}
         \caption{Indoor laboratory layout portraying testbed setup}
         \label{fig:lab}
\end{figure}

\begin{figure}[h]
         \centering
         \includegraphics[width=0.9 \columnwidth]{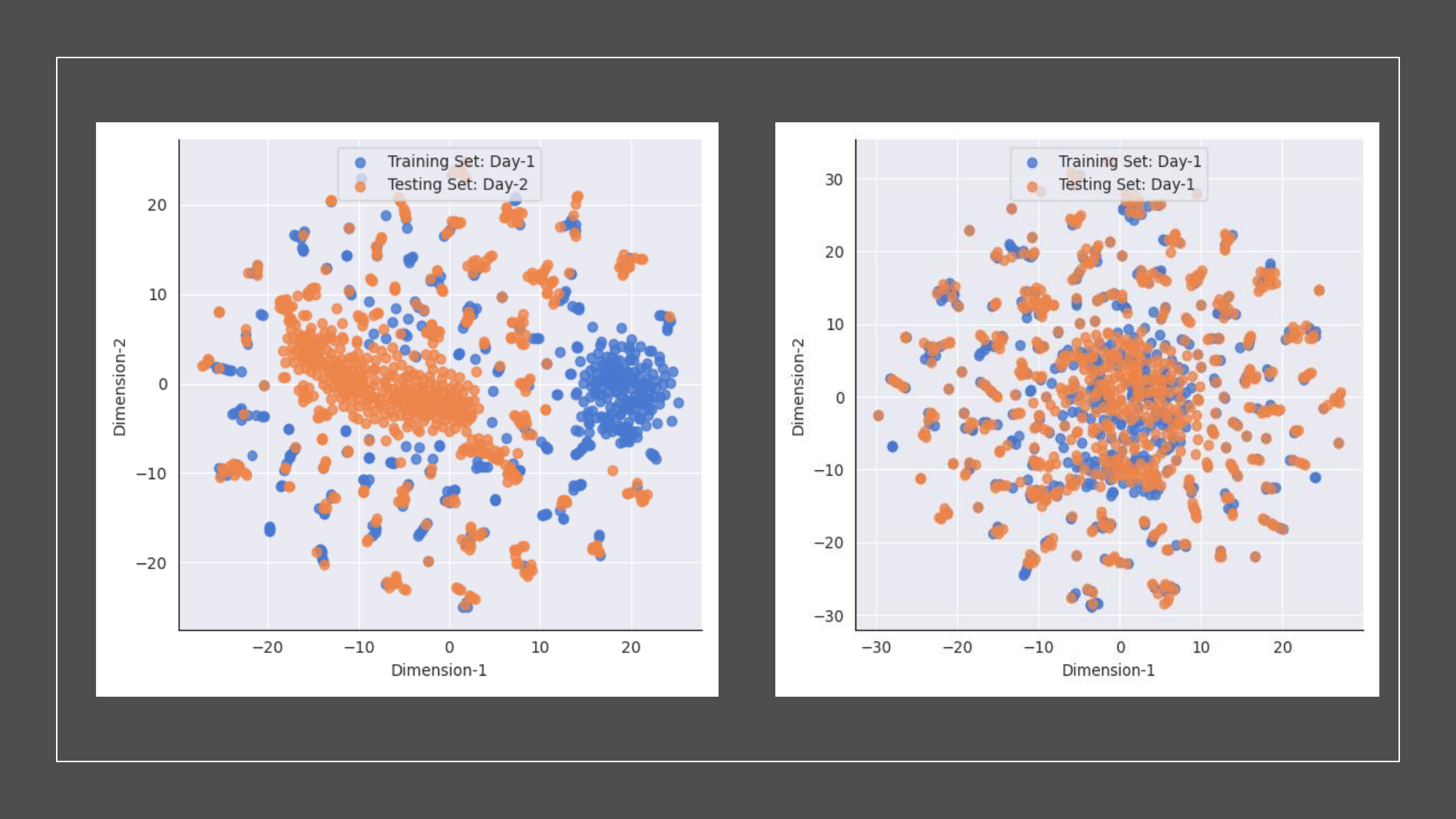}        \caption{t-SNE visualization of training and testing data distribution - Train and test different days (left) - Train and test same day (right).}
         \label{fig:day12}
\end{figure}

\begin{figure}[h]
         \centering
         \includegraphics[width=1 \columnwidth]{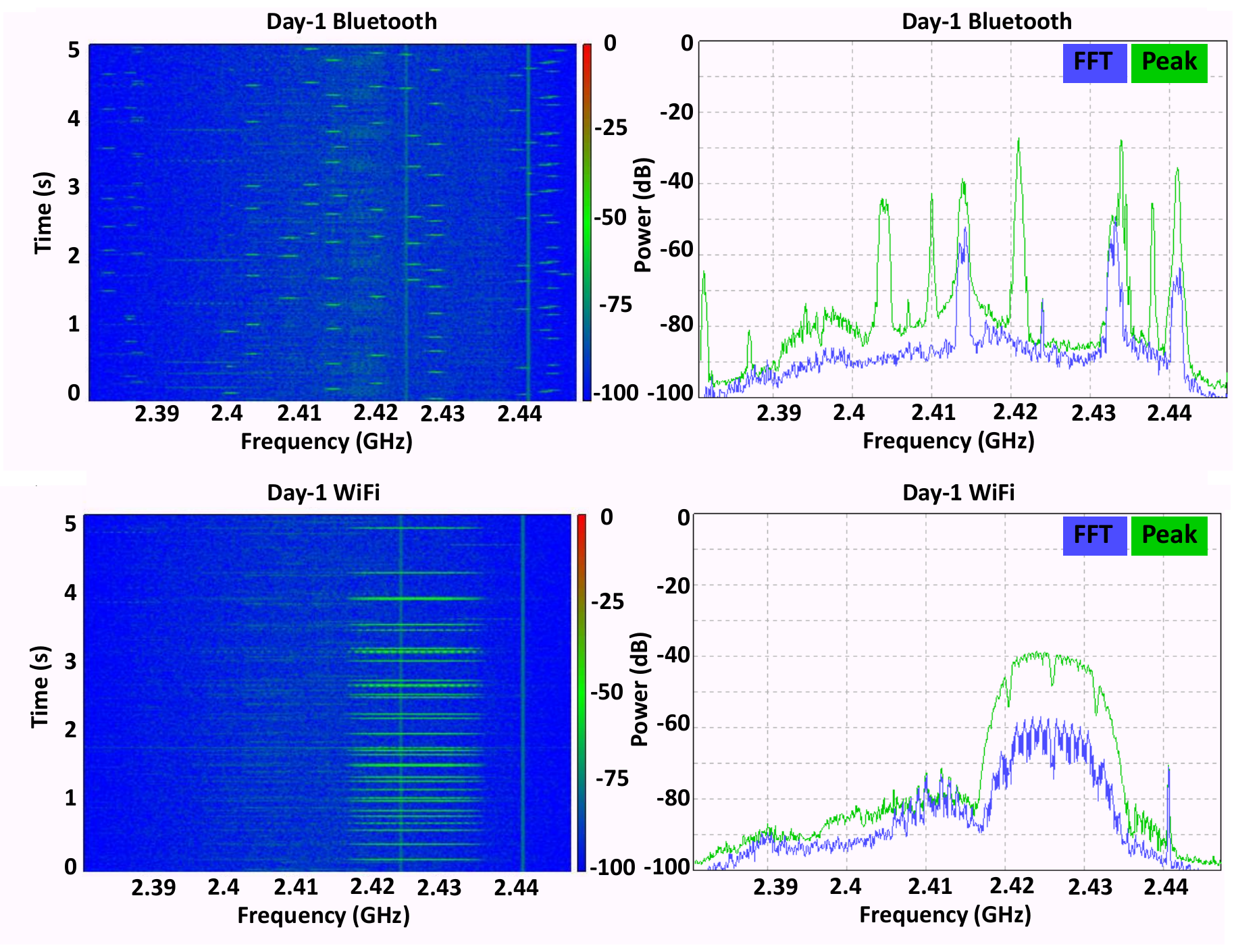}        \caption{Waterfall and PSD of Bluetooth and WiFi captures}
         \label{fig:bt_wifi}
\end{figure}

The experimental testbed for capturing the WiFi and Bluetooth emissions from the IoT devices is shown in Fig.\ref{fig:lab}. The capturing was carried out in an indoor \textit{in-the-wild} laboratory setting with other unavoidable emissions and obstacles pertaining to the indoor setup. The waveforms were collected from one emitter at a time with a passive receiving USRP X300 radio outfitted with a UBX160 daughterboard and VERT2450 antenna in the RX2 antenna port at a sampling rate of 66.67 MS/s with the receiver tuned to 2.44 GHz. We do not use standard WiFi or Bluetooth bandwidth to keep the receiver general and non-specific to the type of emitter. To accomplish this, we use the largest supported sampling rate of the receiver without any sample drops. The emitters used in this data collection are commercial IoT chipsets present in Raspberry Pis and Lenovo laptops. Specifically, we used 2 Lenovo laptops and 8 Raspberry Pi 4Bs. The Raspberry Pis used for the collection possesses a WiFi-Bluetooth combo transceiver chip - Cypress CYW43455 - whereby the WiFi and Bluetooth RF circuitry shares a dual-band antenna. Similarly, the Lenovo laptop uses an Intel Dual Band Wireless-AC 7260 combo chip. Specifically, they generate WiFi IEEE 802.11g and Bluetooth 5.0. To account for signals at varying signal strengths, the distance between the emitter and receiver was varied from 0.5 m to 3 m in steps of 0.5 m corresponding to Bluetooth transmissions. However, the WiFi captures were carried out under varying transmission signal strengths since it was configurable. Owing to the commercial nature of the COTS chipsets, they possessed inherent non-uniformities in the transmission powers. For instance, the Lenovo laptops resulted in a higher power WiFi emission at the same transmission power setting as the Raspberry Pi's Cypress chipset. Hence, the Lenovo laptops were configured to emit lower WiFi transmission power than the Raspberry Pis. Consequently, we had to configure the Raspberry Pis to emit at 5 dBm to 30 dBm in steps of 5 dBm whereas the Lenovo laptops were configured to transmit from 2 dBm to 22 dBm in steps of 4 dBm for the WiFi 802.11g. We would like to note here that the Fig. \ref{fig:lab} is a depiction and does not show all of the emitters and the varying receiver positions to avoid over-crowding and misleading interpretation of the testbed setup. The emitter notation indicates only one emitter was active at a time. Further, in order to capture the device aging with time, the dataset is collected under two different time frames - Day-1 and Day-2. Here the two sets of captures are separated by a period of several months. We validate the argument that the training and testing data under Day-1 and Day-2 have different distributions by utilizing the well-known t-Distributed Stochastic  Neighbor Embedding (t-SNE) visualization tool on the Bluetooth dataset. The training and testing data distribution under the different (left) and same (right) time frame settings are shown in Fig.\ref{fig:day12}. The distribution plot indicates the same distribution for the training and testing set drawn from the same time frame (Day-1) unlike the different time frame (left) where the samples possess different features/distributions. In other words, the plot on the left validates the device aging over the span of several months and that the emitter signature will have slight variations due to device aging. We would like to note here that the aging-related variations in the emitter signature cannot be distinctly captured as the specific receiver circuitry is also subject to such unavoidable aging. In other words, the receiver-recorded samples show a combined effect from the emitter as well as receiver circuitry aging. It is to be noted that the WiFi and Bluetooth emissions are not active simultaneously but only one of them is activated at a time. We used the \textit{rfkill} feature of the chipset to suppress the undesired waveform when the desired waveform standard is active. For example, while collecting Bluetooth emissions we suppressed any undesired WiFi emissions from the same emitter with the command \textit{sudo rfkill block wifi}. The waterfall and PSD plots of WiFi and Bluetooth are shown in Fig.\ref{fig:bt_wifi}.

Each capture within the dataset is long enough (40 MSamples) to support diverse input tensor lengths and formats to suit the user-specific deep learning approach. The dataset also comprises emissions at varying signal powers to account for the feeble to high signal strength emissions as found in a real-world setting. Hence, this dataset amounting to 72 GB of data is intended to foster and accelerate DL-based RF fingerprinting research.

\subsection{Traffic Generation \& Capture}

\subsubsection{Bluetooth}
The Bluetooth 5.0 emissions were generated by initiating a Bluetooth file transfer. Bluetooth 5.0 performs frequency hopping in the 2400 to 2483.5 MHz spectrum and utilizes 40 2 MHz channels. The channelization of carrier frequencies follows the simple math $2402 + k*2 [MHz], \forall k=0,1,...,39$. Bluetooth file transfer was selected since the transfer duration can be controlled by file size. In the setup used for captures, the emitter sends a single large, repetitive file over the air using the OBEX file transfer protocol \cite{OBEXman}. The receiving device was a LG Nexus 5 smartphone running Android version 6.0.1. UNIX tool \emph{hcitool} was used to identify the Bluetooth address of the LG Nexus 5. The hcitool command - \emph{hcitool scan} - makes the Bluetooth device listen for advertisement messages from other Bluetooth devices. This returns the Bluetooth address and common name of any device detected by the scan. The Bluetooth address of the Nexus 5 is required to pair and send files. Device pairing and file transfer were controlled by the command line tool \emph{obexctl}. Obexctl creates an OBEX client and allows device pairing and file transfer from the command line. Pairing is done with command - \textit{connect xx:xx:xx:xx:xx:xx} - 
where \textit{xx:xx:xx:xx:xx:xx} is the Bluetooth address of the Android phone. This command sends a pairing request to the target device (LG Nexus 5). Obexctl prints out \textit{Connection successful} once the pairing is complete. Files can then be sent to the destination with the command - \textit{send file.txt} - to send the text file (file.txt). This command initiates a file transfer to the previously connected device using the file specified. A large text file was selected for transfer to increase transmission duration. This file is 103 MB and contains thousands of lines of identical text. This file transfer had duration of at least 5 minutes, which allowed ample time to capture the signal. 
\subsubsection{WiFi}
In generating the WiFi signal, one desired attribute was that the signal created by the emitter shall be easily detected by the passive listener device. 
The emitters were configured as IEEE802.11g wireless access points (WAPs) to broadcast a beacon message at a defined interval and channel. By setting the WiFi channel and beacon interval, the emitted signal is more receptive to capture. The emitter utilized the WiFi channel 8 centered at 2447 MHz and a 20 MHz bandwidth spanning 2436-2458 MHz. This is indicated in the host access point daemon file as hw\_mode$=$g and channel$=$8 corresponding to WiFi standard (IEEE802.11g) and channel specification respectively. When the emitter device is configured as a WAP, the emitter will begin broadcasting beacon messages immediately. This can lead to undesired transmissions if multiple emitters are powered ON. To prevent simultaneous transmissions, \emph{rfkill} is used to enable/disable the WiFi on the emitter before and after captures. WiFi beacon generation is started by unblocking WiFi with the command - \textit{rfkill unblock wifi}. Conversely, to prevent WiFi emissions, WiFi is blocked with the command - \textit{rfkill block wifi}.
\subsubsection{Firmware}    

The \textit{host access point daemon }is used to convert the emitter to a WAP. A configuration file on the emitter specifies the wireless network parameters. 
WiFi channel 8 was chosen because the center frequency (2447 MHz) is near the center of the 2.4 GHz ISM band. The beacon interval is the lowest possible value at 15 milliseconds. This will produce a periodic signal at center frequency 2447 MHz with a transmit frequency of 66.67 signals per second. 

Moreover, \emph{dnsmasq} is required on the emitting device to properly create an access point. dnsmasq provides services needed by the access point such as DNS caching, DHCP server handling, and router advertisement. The router advertisement messages are the backbone of the WiFi beacon messages. the dnsmasq is configured in config file \emph{/etc/dnsmasq.conf}. Fields required in \textit{dnsmasq.conf} are the wireless interface, the router IP address, and a range of IP addresses to be used by the DHCP server.  This file is identical for the Laptops except the wireless interface is changed to the laptops' wireless interface. 
Additionally, \emph{dhcpcd} is used to give the emitter device a static IP address. The static IP address is required by the DHCP client configured in dnsmasq.conf. The dhcpcd is configured in a config file \emph{/etc/dhcpcd.conf}. A few lines were added to the default \textit{dhcpcd.conf} file to give a static IP address to the wireless interface. 
These configuration lines specify the network interface \textit{wlan} shall use static IP address - 192.168.4.1 - which is the same value specified in the dnsmasq config file. The line \textit{nohook wpa\_supplicant}  disables the WPA supplicant service. WPA supplicant must be disabled to configure the emitter as a WAP.   

\subsection{Dataset Organization}
Each dataset is organized into a collection of raw binary captures each associated with its corresponding JSON metadata file. Consequently, for each binary file containing the raw IQ samples, there is a JSON metadata file under the same filename. The dataset is a modified version of the SigMF format with field extensions to suit our applications. Table \ref{tab:meta} shows the various metadata fields in the JSON file that helps record every aspect of the data collection. Each metadata field is stored as a key-value pair. Such a dataset organization helps mimic the experimental testbed setup to facilitate reproducibility. Since we capture six different transmitter-receiver separations (signal strengths) for each emitter, each dataset possesses 60 captures in total from the 10 IoT devices. Hence, the Day-1 and Day-2 datasets together possess 120 captures in raw binary format and their corresponding JSON metadata files. Therefore, the WiFi and Bluetooth captures possess 120 captures each. Each binary capture file records 40 MSamples at a sampling rate of 66.667 MS/s with the receiver tuned in at 2.44 GHz. 

\begin{table*}
\centering
\newcommand*{\MyIndent}{\hspace*{0.8cm}}
\caption{Metadata Structure in the JSON file \label{tab:meta}}
\begin{tcolorbox}[width=16.0 cm,tab2,tabularx={p{2.4 cm}|p{3.3 cm}|p{9 cm}}]

\textbf{Top-level JSON Objects} & \textbf{Metadata Fields} & \textbf{Description} \\
\hline
\textbf{global} &version &	The version of the SigMF standard referenced. \\
& sample\_rate &The sampling rate of the waveform being recorded. \\
& total\_transmissions &The total emissions in this dataset. \\
& description &A short description of the contents of this file and its corresponding binary capture file.\\
& record\_date &The date on which this capture was recorded.\\
& datatype &The type of data in the binary capture file, e.g., float16, float32, complex64, etc.\\\hline

\textbf{captures} &center\_frequency & The center frequency of the recording in Hz. \\
& sample\_start & The index indicating the start of waveform samples in the recording.
\\
& transmission\_number &The index of the recorded transmission. \\\hline
\textbf{annotations} &environment &The environment where the waveform is being recorded. \\
& distance &The distance between the emitter and the passive receiver. \\
&sample\_count & The total number of samples in the recording. \\
&protocol &The wireless standard of the waveform being recorded.\\ 
 &\textbf{transmitter}\\     &\MyIndent antenna &The antenna used by the emitter. \\
 &\MyIndent antenna\_port &The port where the transmitter antenna is connected. \\
 &\MyIndent radio\_id &The serial number to identify the emitter radio.\\
 &\MyIndent tx\_mode &The mode of transmission - single or multiple.\\
 &\MyIndent modulation &The modulation format of the waveform being recorded. \\
 &\MyIndent medium &The propagation medium of the emitted waveform. \\
 &\MyIndent gain &The transmission gain of the emitted waveform. \\
 &\MyIndent make\_and\_model &The make and model of the emitter radio. \\
 &\MyIndent daughter\_board &The daughterboard of the emitter radio. \\
 &\textbf{receiver}\\     &\MyIndent antenna &The antenna used by the receiver.  \\
 &\MyIndent antenna\_port &The port where the receiver antenna is connected. \\
 &\MyIndent radio\_id &The serial number to identify the receiver radio.\\
 &\MyIndent radio\_ip\_addr &The IP address to identify the receiver radio.\\
 &\MyIndent radio &The receiver radio.\\
 &\MyIndent rx\_mode &The mode of reception - single or multiple antennae.\\
 &\MyIndent medium &The propagation medium of the emitted waveform.\\
 &\MyIndent gain &The receiver gain of the passive receiver. \\
 &\MyIndent make\_and\_model &The make and model of the receiver radio. \\
 &\MyIndent daughter\_board &The daughterboard of the receiver radio. \\
 &\MyIndent samp\_rate &The sampling rate of the receiver radio. \\
 &\MyIndent filename &The filename of the binary file containing the recorded samples. \\
\hline
\end{tcolorbox}
\end{table*}
\subsection{Utilizing the Dataset - Parsing Dataset Tutorial}

The recorded samples can be effortlessly read from the binary (.dat) file while extracting the metadata from the JSON file. The example Python snippet shared in Code Ocean \cite{code} shows how the various metadata fields can be extracted from the JSON file.

\section{Research Use Case Example}
The introduced dataset supports different DNN architectures of low to high density due to its ability to support very large number of training examples of multiple lengths. This will be extremely useful for novel RF fingerprinting solutions that are designed for IoT security schemes such as device authentication, intruder detection, compliance violation detection, user-specific traffic monitoring, among others. The dataset will support RF fingerprinting solutions tailored towards lightweight edge deployment as well as deployment on extremely powerful GPU servers. For example, lightweight edge deployment would be where the power, memory, and computational resources are limited and requires lightweight neural networks. Such neural networks may require training on shorter bursts of data samples which can be easily acquired from the extremely large capture provided in each capture file. On the other hand, if the DNN models require observation of longer capture lengths over several thousands of samples, then that will also be supported by the provided capture length ($\sim40$ MSamples). Such long captures will aid diverse RF fingerprinting approaches involving user-specific preprocessing and user-defined lengths. For example, some approaches might specifically process the received waveform to select only the burst present regions of the capture of specific lengths (128, 1024, 128000, etc.) and may be preprocessed to obtain the magnitude, phase, or time-frequency energy distribution. Since the capture is available as a large file of complex64 IQ samples that are in its raw unprocessed form, it will support such custom use cases. 

This dataset was used in our previous work \cite{AJagannath22GLOBECOM} where we used the dataset for single-task and multi-task model evaluations under single and mixed-day scenarios. Here, we chose the burst present regions from the captures of length 1024. The bursts of desired lengths were chosen from the capture files of WiFi as well as Bluetooth waveforms. Since the dataset provides access to raw complex IQ samples any custom preprocessing or other transformations inherent to the custom DNN models can be directly applied to the samples. For example, the architecture in \cite{AJagannath22GLOBECOM} had a time-frequency transformation branch within its novel attentional framework. This was only possible since the samples are raw IQ values. The different time frame captures were utilized to perform single and mixed days dataset distribution to evaluate the model's performance when trained and tested with samples drawn from single and mixed time frames distribution respectively. We deduced that the mixed-days dataset yielded better accuracy since it accounted for a much more diverse dataset distribution under the single-task setting. It was also concluded that to enhance Bluetooth fingerprinting accuracy, the input sample lengths must be longer than 1024. Such higher lengths are also possible with the released dataset since the capture files are very large (each comprising 40 MSamples). Additionally, if the user requires hand-crafted features as input to their DNN models, say, for example, short-time Fourier transformed tensors, custom data normalization on shorter burst lengths, magnitude, and phase among others, these can be easily obtained from the raw IQ values instead of already transformed and scaled values. Therefore the utility of this dataset is manifold. 


\section{Conclusion}
We introduced a real-world and diverse RF dataset comprising emissions from COTS IoT chipsets. Most importantly, we share with the community a unique dataset containing heterogeneous wireless standards emitted from the same antenna of WiFi-Bluetooth combo chips. If a DNN is trained only on a single wireless standard say WiFi, it might not effectively identify such combo chips when it transmits Bluetooth waveform. With this dataset, we aim to bridge this existing deficiency and share with the public a first-of-its-kind RF fingerprinting dataset. Additionally, to support generalization tests, we share WiFi and Bluetooth datasets collected over two different time frames. Finally, the modified SigMF format of the dataset will facilitate seamless reproducibility of the data collection environment.

\bibliography{bibfile.bib}

\bibliographystyle{IEEEtran}

%



\begin{IEEEbiography}[{\includegraphics[width=1in,height=1.25in,clip,keepaspectratio]{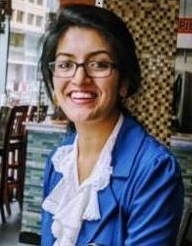}}] {Anu Jagannath}[SM'21] received the M.S degree from State University of New York at Buffalo in 2013. She is the Associate Director of Marconi-Rosenblatt AI/ML Innovation Lab at ANDRO Computational Solutions, LLC. She is also pursuing her Ph.D. at Northeastern University. Her research interests are dynamic signal intelligence, extremely compact neural network architectures for edge deployment, adaptive signal processing, MIMO, Machine learning, cognitive networking, adaptive transceivers, among others. She has authored multiple peer-reviewed conference, journal articles, and book chapter while also rendering her reviewing services to peer-reviewed journals and conferences. She has served on the Technical Program Committee of IEEE ICC Machine Learning for Communications Track for two consecutive years (2022 and 2023). She has been the Co-PI of several SBIR efforts for the US Army, US SOCOM, and DHS. Her research has led to 6 US patents (granted and pending).
\end{IEEEbiography}
\vskip -2.5\baselineskip 

\begin{IEEEbiography}[{\includegraphics[width=1in,height=1.5in,keepaspectratio]{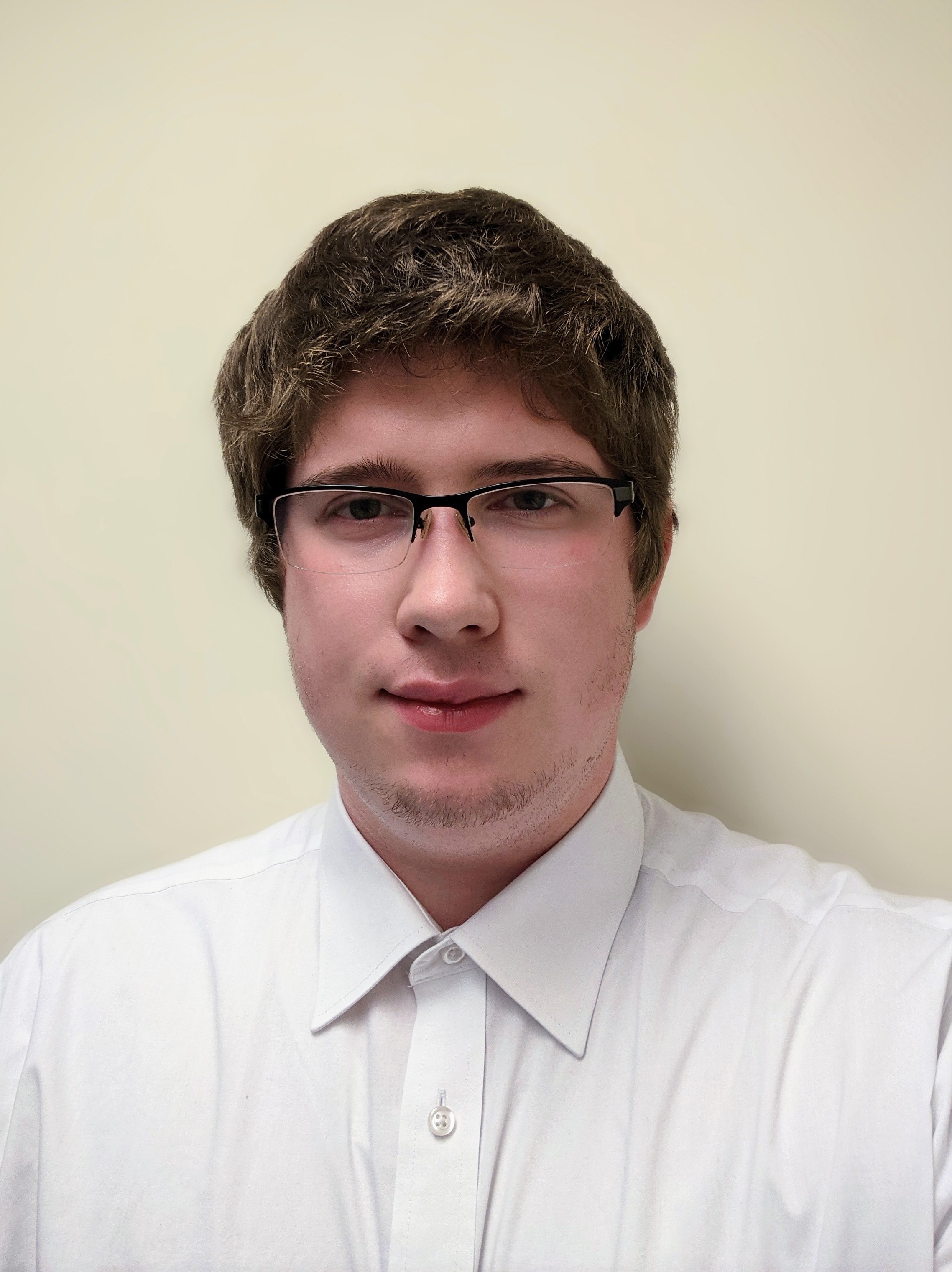}}]{Zackary Kane} received a B.S. in Electrical and Computer Engineer from the State University of New York Polytechnic Institute. Zack is currently a Radio Frequency Engineer/Associate Scientist I at ANDRO Computational Solutions in ANDRO's Marconi-Rosenblatt lab. He has worked on two Small Business Innovation Research (SBIR) projects and two Rapid Innovation Fund (RIF) projects. In the past, Zack assisted in range testing the SPEARLink\texttrademark~radio platform and worked on documentation for the SPEARLink\texttrademark~radio. He is currently assisting in the development of a multi-band signal classification model. 
\end{IEEEbiography}

\begin{IEEEbiography}[{\includegraphics[width=1in,height=1.25in,clip,keepaspectratio]{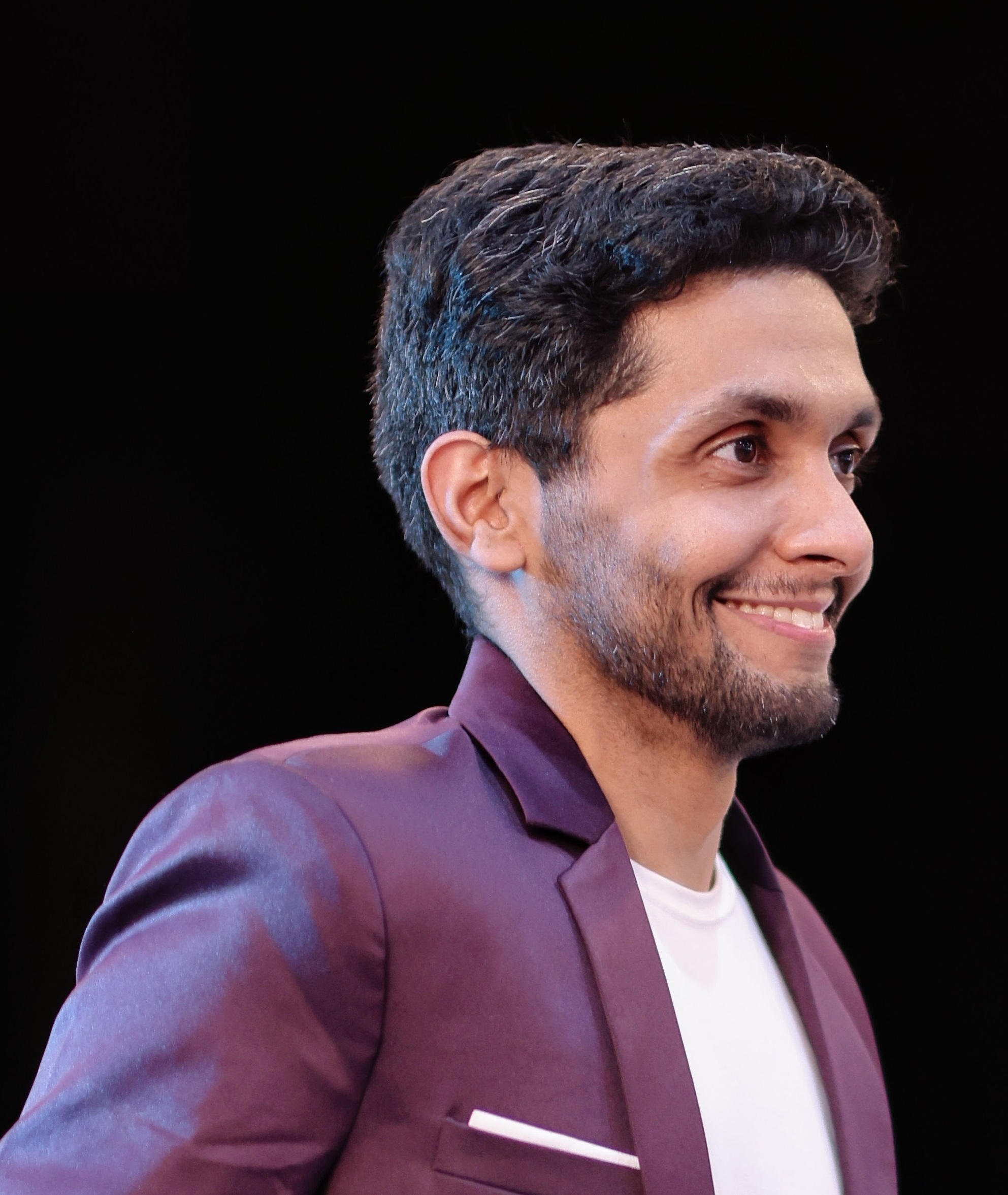}}]
 {Jithin Jagannath}[SM'19] received Ph.D. degree from Northeastern University. He is the Founding Director of ANDRO's Marconi-Rosenblatt AI/ML Innovation Lab and Adjunct Assistant Professor at the University at Buffalo. He has been the Principal Investigator of several cutting-edge research projects exploring radio Frequency machine learning (RFML), RF SIGINT, and autonomous UAS for U.S. Army, NAVAIR, SOCOM, AFSOR, ONR, and DHS. He has been invited to give Keynotes and as a Panelist to discuss his vision on applied machine learning, Beyond 5G, and RF SIGINT at leading IEEE Conferences. He serves on the editorial board of Computer Networks (Elsevier) and TPC for leading IEEE Conferences Tracks related to RFML. His research has led to over 40 peer-reviewed publications and 15 patents (granted and pending). Dr. Jagannath was the recipient of the 2021 IEEE Region 1 Technological Innovation Award. He is also the recipient of AFCEA International Meritorious Rising Star Award and AFCEA 40 Under Forty.  
\end{IEEEbiography}

\newpage

\end{document}